\documentclass[aps,prl,superscriptaddress,showpacs,twocolumn,floatfix]{revtex4}
\usepackage{graphicx}
\usepackage[usenames,dvipsnames]{color}

\newcommand{\etal}{{\em et al.}}
\newcommand\JLab{Thomas Jefferson National Accelerator Facility, Newport News, VA 23606 }
\begin{document}
\preprint{APS/123-QED}
\title{On measurement of the isotropy of the speed of light}
\author{B.~Wojtsekhowski}
\affiliation{\JLab}
\date{\today}

\begin{abstract}
Three experimental concepts investigating possible anisotropy of the speed of light are presented. 
They are based on i) beam deflection in a 180$^\circ$  magnetic arc,
ii) narrow resonance production in an electron-positron collider, and 
iii) the ratio of magnetic moments of an electron and a positron moving in opposite directions.
\end{abstract}


\maketitle

There are several well known experiments which investigate the one-way speed of light $c$; 
see the review and analysis in Refs.~\cite{CW1992, CW2014}.
Here we present related experimental schemes based on a high energy electron (positron) beam.
The experiments test isotropy of the maximum speed, but we refer to the speed of light 
assuming the photon to be massless.
A recent experiment at the storage ring ESRF (Grenoble) established a stringent constraint 
on the speed of light anisotropy of $1.6 \times 10^{-14}$ using the Compton 
back-scattering method~\cite{GM1996, GR2010}.

In this letter we discuss three other methods which could be used in a search for 
directional variation of the speed of light and/or related effects.
Their common feature is a large value of the Lorentz factor of the electron (positron) 
beam, $\gamma_{-(+)}$, of a few $10^4$.
They are based on various techniques for the beam momentum measurement.

\section{Momenta in opposite directions} 
\begin{figure}[thb]
\centering
\includegraphics[width=0.25\textwidth]{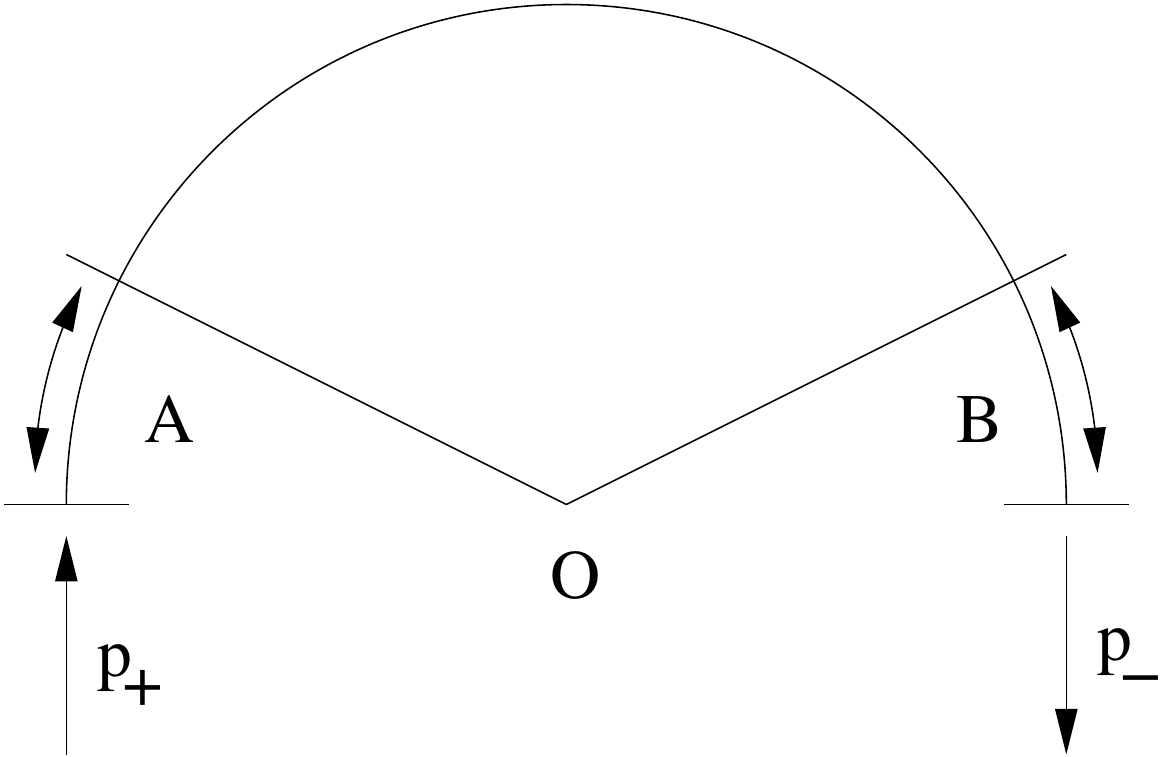} 
\caption{Diagram for measurement of the beam momenta with a 180$^\circ$ arc magnet.}  
\label{fig:arc}
\end{figure}

Deflection of the beam in a magnetic field is the simplest method for a momentum measurement. 
Such a measurement could provide very accurate monitoring of slight changes of momentum 
assuming stability of the magnetic field and the beam position detectors.
Two beam momentum monitors at opposite sides of a 180$^\circ$ bending magnet allow one
to measure the ratio of the particle momenta $p_+, \, p_-$ moving in opposite directions, $R = p_+ / p_-$ (see 
Fig.~\ref{fig:arc}).
The sensitivity of the method, $\delta c/c \sim (\delta R/R) /\gamma^2$, is defined by 
the beam Lorentz factor and relative accuracy of the momentum monitors.

Assuming that there are no acceleration elements between those two momentum 
measurements and a small correction (calculable) on the radiative energy loss, 
such a ratio should be stable even when the actual beam energy varies.
The value of $R$ will be close to 1 with any deviation mainly due to the systematics 
uncertainties of the momentum monitors.
The configuration could have a large portion of the arc used as a momentum monitor
(the area A (B) shown in Fig.~\ref{fig:arc}) because it allows higher dispersion 
and better sensitivity in spite of the angle's being smaller than 180$^\circ$ 
between the two momentum monitors.
The systematics of the $R$ measurement are strongly suppressed when the measurements in 
both monitors perform synchronously.
The proposed measurement of the ratio $R$ is sensitive to the directional variation of 
the speed of light as formulated, for example, in the Mansouri-Sexl test theory~\cite{TestT}.
The results could also be interpreted in terms of the Standard Model Extension~\cite{SME}, 
whose odd-parity parameter $\tilde{k}_{o \,+}$ would be constrained.
The time dependence of $R$ could reveal, for example, the directional variation of the speed of light 
due to changes in the beam direction because of the accelerator rotation together with 
the Earth and variation of the speed of motion relative to 
the Cosmic Microwave Background dipole~\cite{CMB}.

A number of accelerators with beam energies in the several GeV range have (or could build)
beam momentum monitors with a level of $10^{-6}$ relative precision over a period of one millisecond. 
Indeed, the accuracy of the momentum monitor is defined by the precision of the beam position monitors, 
BPMs, and the deflection of the beam trajectory in the dipole magnet. 
The RF-based BPMs have an accuracy on the level of 1~$\mu$m for a 0.1 nC beam bunch charge
~\cite{KEK-BPM}.
Measurement of the magnet temperature with 0.01$^\circ$K accuracy and the magnetic field by 
means of NMR should allow sufficient short term stability of the magnetic field integral.
The BMPs located next to the dipole section with dispersion of 2-3 meters should provide the required
accuracy of the momentum measurement.
When data are averaged over a few hours, the potential sensitivity of $R$ should be
on the level of $10^{-9}$ (limited by the long term stability of the beam position monitors, 
which could be determined by means of the optical interferometer to sufficient precision
as shown e.g. in Ref.~\cite{ILC-BPM}).
The relatively small size of each momentum monitor of 20 meters 
will help to minimize known distortions such as tide and ground instability.
The combined effect of these uncertainties on the precision of the $R$ measurement 
we estimate to be less than $10^{-7}$.

The stability of measurement could be greatly improved by means of two beams 
(electron and positron) moving in the same magnetic system, which would allow a full
compensation of drift of the  momentum monitor characteristics with normalization
of the electron ratio $R^e = p_+^e / p_-^e$ to the positron ratio $R^p = p_+^p / p_-^p$.
The double ratio $R_{e/p} = R^e/R^p$ would be immune to most instabilities.
This double ratio has also doubled sensitivity to the directional variation of 
the speed of light.
Such a measurement could be performed with a single storage ring collider, e.g. 
CESR or VEPP-4.

The sidereal periodicity of the possible signal when used in
the Fourier analysis of the data will allow suppression of the systematics by a factor of 10-20
depending on the duration of the experiment.
By combining the resulting sensitivity of the momenta ratio measurement $R$ of $10^{-8}$
with the factor of the beam ($\gamma^{-2} \sim 10^{-8}$), one can find that the
signal sensitivity is $10^{-16}$ or better for $\delta c/c$, 
which would be a significant improvement for the limit on the odd-parity anisotropy,
whose current value is $10^{-14}$~\cite{CAV}.

The JLab CEBAF accelerator~\cite{CEBAF} has the most advantageous parameters 
for the proposed experiment because its beam has a high current of 100 $\mu$A,
a relative energy spread of a few 10$^{-5}$, and 
geometrical emittance of 10$^{-9}$ m$\times$rad (in spite of significant broadening
due to radiation losses in the last few turns).
The accelerator sections are separated by 180$^\circ$ arcs, 
and the beam recently reached $\gamma \sim 2 \times 10^4$,
the highest Lorentz factor among the currently operating accelerators. 
The absolute value of the beam energy in a linear accelerator 
could vary, but this does not impact the proposed investigation because only the ratio
of the beam momenta at opposite sides of the arc needs to be stable.
In addition, the value of the anisotropy could be constrained
at 10 different arcs with increasing beam energy.

\section{Resonance production}

The masses of the narrow resonances ($\varphi, J/\Psi, \Upsilon $, and Z) 
have been measured with high precision in electron-positron colliders.
For example, the mass of the $J/\Psi$ is known with a relative accuracy of $3.5 \times 10^{-6}$
and the mass of the $\Upsilon$ with $1 \times 10^{-5}$~\cite{PDG}.
These collider results were obtained using a precision measurement of the beam energies via
the spin resonance method, which allows accurate determination of the absolute value of the
beam Lorentz factor~\cite{POL}.
Variation of the speed of light leads to a change of the Lorentz factor along the beam orbit
in a storage ring. 
The resonance could be used as the second method of beam energy measurement
needed in a search for a sidereal time variation of the beam $\gamma$.

Because the beam directions at the location of the energy meter and the $e^+e^-$ collision point are
different (in some experiments), the average observed Lorentz factor of the beams 
is not necessarily equal to the ratio of the meson and electron masses.
For a symmetric-energy collider (equal energies of a positron and an electron beam)
to first order the effect is canceled out.
The B-factories at SLAC and KEK~\cite{BFAC} could access the non-isotropic 
component of the speed of light due to the large difference in the Lorentz factor of the electron and 
positron beams.
The variation in the ratio $\beta = v/c$ could be expressed as 
${\delta \beta}/{\beta} \,=\, {\delta m}/{m} /(\gamma_+^2 - \gamma_-^2)$.
The full observed width of the $\Upsilon$ resonance is of 10 MeV (mainly
due to the beam energy spread).
This would lead to an estimate of a potential constraint on $\delta c/c$ of $3 \times 10^{-12}$.
Finally, very large statistics of accumulated $\Upsilon$ events in the
KEK experiment~\cite{Belle}, $1 \times 10^8$, boosts potential sensitivity to $3 \times 10^{-16}$.
The sidereal variation of the observed resonance mass (at the fixed beams energies) would be a signal
of the anisotropy of the speed of light.

\section{Leptons' anomalous magnetic moments}
The anomalous magnetic moments $a_{-(+)}$ of an electron and positron were measured to a
$2 \times 10^{-12}$ relative accuracy in single particle traps~\cite{g-2t}, which puts
a stringent constraint on the CPT violating parameter of SME~\cite{SME}.
Independent measurement on the level of $ 10^{-8}$ was performed
at the VEPP2m storage ring with 650~MeV beams of electrons and positrons~\cite{g-2s}.
These beams of electrons and positrons move in the storage ring in opposite directions
and their Lorentz factors are obtained by the spin resonance method.
Assuming that $a_+ = a_-$ (constraint~\cite{g-2t}) and equal masses 
of an electron and a positron, we can find the difference of the Lorentz factors of these
two beams and use such an experiment for a search of the anisotropy of the speed of light.
Sensitivity of the experiment~\cite{g-2s} corresponds to \mbox{$\delta c/c \sim$ 10$^{-14}$}.

There is the potential to reach a higher precision of $(a_+ - a_-)/a_{avg}$ of $10^{-10}$ 
with the beam method~\cite{POL};
alternatively, one can perform such a measurement using the VEPP-4 storage ring,
where an almost ten times higher beam $\gamma$ factor leads to higher \mbox{$\delta c/c$} sensitivity.
The resulting estimate for $(\gamma_+ - \gamma_-)/\gamma_{avg}$ of $10^{-10}$ means that 
the search sensitivity for \mbox{$\delta c/c$} is about $10^{-16}$
from multiple measurements with different sidereal phases.

\acknowledgments
The author takes pleasure in acknowledging helpful discussions with V. G.~Gurzadyan 
of the GRAAL experiment and with C.~Keppel, B.~Vlahovic, and V.~Zelevinsky of the current concepts. 
He would like to extend thanks to P.~Evtushenko and Y.~Roblin for information on CEBAF 
parameters. This work was supported by the U.S. Department of Energy. 
Jefferson Science Associates, LLC, operates Jefferson Lab for 
the U.S. DOE under U.S. DOE contract DE-AC05-060R23177.

\end{document}